# Directed and Elliptic Flow in 158 GeV/Nucleon Pb + Pb Collisions

Draft 2.2

H. Appelshäuser (7,#), J. Bächler(5), S.J. Bailey(16), L.S. Barnby(3), J. Bartke(6),
R.A. Barton(3), H. Bialkowska(14), C.O. Blyth(3), R. Bock(7), C. Bormann(10), F.P. Brady(8),
R. Brockmann(7,±), N. Buncic(5, 10), P. Buncic(5, 10), H.L. Caines(3), D. Cebra(8),
G.E. Cooper(2), J.G. Cramer(16), P. Csato(4), J. Dunn(8), V. Eckardt(13), F. Eckhardt(12),
M.I. Ferguson(5), H.G. Fischer(5), D. Flierl(10), Z. Fodor(4), P. Foka(7, 10), P. Freund(13),
V. Friese(12), M. Fuchs(10), F. Gabler(10), J. Gal(4), M. Gazdzicki(10), E. Gladysz(6),
J. Grebieszkow(15), J. Günther(10), J.W. Harris(17), S. Hegyi(4), T. Henkel(12), L.A. Hill(3),
I. Huang(2, 8), H. Hümmler(10,+), G. Igo(11), D. Irmscher(2, 7), P. Jacobs(2), P.G. Jones(3),
K. Kadija(18, 13), V.I. Kolesnikov(9), M. Kowalski(6), B. Lasiuk(11, 17), P. Levai(4),
A.I. Malakhov(9), S. Margetis(2,*), C. Markert(7), G.L. Melkumov(9), A. Mock(13),
J. Molnar(4), J.M. Nelson(3), G. Odyniec(2), G. Palla(4), A.D. Panagiotou(1), A. Petridis(1),
A. Piper(12), R.J. Porter(2), A.M. Poskanzer(2), S. Poziombka(10), D.J. Prindle(16),
F. Pühlhofer(12), W. Rauch(13), J.G. Reid(16), R. Renfordt(10), W. Retyk(15), H.G. Ritter(2),
D. Röhrich(10), C. Roland(7), G. Roland(10), H. Rudolph(2, 10), A. Rybicki(6),
A. Sandoval(7), H. Sann(7), A.Yu. Semenov(9), E. Schäfer(13), D. Schmischke(10),
N. Schmitz(13), S. Schönfelder(13), P. Seyboth(13), J. Seyerlein(13), F. Sikler(4),
E. Skrzypczak(15), G.T.A. Squier(3), R. Stock(10), H. Ströbele(10), I. Szentpetery(4),
J. Sziklai(4), M. Toy(2, 11), T.A. Trainor(16), S. Trentalange(11), T. Ullrich(17),
M. Vassiliou(1), G. Vesztergombi(4), S. Voloshin(2,#, &), D. Vranic(5, 18), F. Wang(2),
D.D. Weerasundara(16), S. Wenig(5), C. Whitten(11), T. Wienold(2,#), L. Wood(8),
T.A. Yates(3), J. Zimanyi(4), R. Zybert(3)

(NA49 Collaboration)

(1)Department of Physics, University of Athens, Athens, Greece.
(2)Lawrence Berkeley National Laboratory, University of California, Berkeley, CA, USA.
(3)Birmingham University, Birmingham, England.
(4)KFKI Research Institute for Particle and Nuclear Physics, Budapest, Hungary.
(5)CERN, Geneva, Switzerland.
(6)Institute of Nuclear Physics, Cracow, Poland.
(7)Gesellschaft für Schwerionenforschung (GSI), Darmstadt, Germany.
(8)University of California at Davis, Davis, CA, USA.
(9)Joint Institute for Nuclear Research, Dubna, Russia.
(10)Fachbereich Physik der Universität, Frankfurt, Germany.
(11)University of California at Los Angeles, Los Angeles, CA, USA.
(12)Fachbereich Physik der Universität, Marburg, Germany.
(13)Max-Planck-Institut für Physik, Munich, Germany.
(14)Institute for Nuclear Studies, Warsaw, Poland.
(15)Institute for Experimental Physics, University of Warsaw, Warsaw, Poland.
(16)Nuclear Physics Laboratory, University of Washington, Seattle, WA, USA.
(17)Yale University, New Haven, CT, USA.
(18)Rudjer Boskovic Institute, Zagreb, Croatia.

The directed and elliptic flow of protons and charged pions has been observed from the semi-central collisions of a 158 GeV/nucleon Pb beam with a Pb target. The rapidity and transverse momentum dependence of the flow has been measured. The directed flow of the pions is opposite to that of the protons but both exhibit negative flow at low $p_t$. The elliptic flow of both is fairly independent of rapidity but rises with $p_t$.

The azimuthal anisotropy of charged particle emission from the interaction of a 158 GeV/nucleon Pb beam with a Pb target has been studied in the two main Time Projection Chambers (TPCs) of CERN SPS experiment NA49.[1] The TPCs, situated downstream of two large dipole magnets, cover a large region of phase space forward of mid-rapidity. Identification of protons and pions is performed by measurements of energy loss of the detected particles in the gas of the TPCs. The large phase-space acceptance allows event-by-event study of the angular correlations of the particles from the interaction. It is thought that angular correlations generated by collective flow retain some signature of the effective pressure achieved at maximum compression in the interaction.[2,3] This is the first study of directed and elliptic flow as a function of rapidity and transverse momentum for collisions of the heaviest nuclei at the highest bombarding energy presently available.

Usually three kinds of flow in the plane transverse to the beam are considered: radial transverse flow, directed flow, and elliptic flow. For central collisions which are azimuthally isotropic, only radial transverse flow is allowed. One of its effects is to raise the apparent temperature of the single-particle transverse momentum spectra. For non-central collisions, a plane can be determined for each event describing the azimuthal anisotropy of the event, and directed and elliptic flow can be identified from the azimuthal anisotropy of the particles with respect to this plane. In a Fourier expansion[4-6] of the azimuthal distribution of the particles with respect to this plane the amplitude of the first harmonic of the distribution corresponds to the directed flow which was discovered at the Bevalac.[7] One of the measures of directed flow, the mean momentum in the flow direction, appears to peak at beam energies of about one GeV/nucleon and then decreases at higher energies.[8] Except at the very lowest beam energies, the directed flow of the protons is thought to be on the side of the beam away from the target nucleus. On the other hand, the directed flow of the produced pions is often opposite to that of the protons because of shadowing effects. For a recent review of flow see ref. [9]. From the data presented here the decrease from the peak values at Bevalac-SIS energies to SPS energies is a factor of about 400 for the mean polar flow angle and a factor of 5-10 for the mean momentum in the flow direction. Thus, directed flow is a much smaller effect at the SPS then at lower beam energies. The amplitude of the second harmonic of the azimuthal distribution of particles with respect to the event plane measures the elliptic flow, the importance of which at high energies was first emphasized by Ollitrault.[10] At Bevalac energies elliptic flow for both protons[5] and produced pions[11] was found to be oriented perpendicular to the directed flow and was called squeeze-out, but at high energies elliptic flow is expected to be in the plane of the directed flow[3, 10, 12], and this has been found at the AGS.[13] From the data presented here it will be shown that the elliptic flow at the SPS is comparable to that at Bevalac-SIS energies but oriented in the plane of the directed flow instead of perpendicular to the plane as at Bevalac energies.

The data presented here consist of 50k events taken with a medium bias trigger as determined by the NA49 veto calorimeter which measures the energy within 0.3° of the beam. This trigger selected events with veto calorimeter energy[14] from 0.45 to 0.6 of the beam energy, and corresponds to an impact parameter selection of about 6.5 to 8.0 fm, as estimated from VENUS[15] simulations.

Particle identification was performed in the TPCs by measuring the specific energy loss in the gas and was used to identify highly enriched samples of protons and charged pions. The proton sample used in this analysis had laboratory momenta greater than 30 GeV/c. It had an observed rapidity distribution peaked between 4 and 5.25 and an observed mean multiplicity of about 20. By comparison with the yield of negative particles in the same energy-loss window it was estimated that this proton sample was enriched to about 85% in protons. After removing this proton sample, the other positive and negative charged particles formed a sample called charged particles. However, the particles in the proton energy-loss window between 10 and 30 GeV/c were not included in either the proton sample or this charged particle sample. From this charged particle sample, those particles with rapidity (assuming the pion mass) from 4 to 6, and transverse momenta from 0.05 to 1.0 GeV/c were used to determine the orientation of the event plane. They had a mean observed multiplicity of about 170. They were also used for the results which are integrated over rapidity and $p_t$. The sample identified as pions had momenta between 3 and 50 GeV/c and was thought to be highly enriched in pions based on fits of four Gaussian distributions (p, K, π, e) to the energy-loss spectra. This pion sample had an observed rapidity distribution peaked around 4 and a observed mean multiplicity of about 120. This pion sample was used for the results to be presented as a function of rapidity and $p_t$.

Both first harmonic and second harmonic event planes, called here respectively the plane and the ellipse, were determined event by event. The azimuthal laboratory angles of these planes were calculated with the following equation:

$$\emptyset_n = \left( \tan^{-1} \frac{\sum wgt(\emptyset_i) \sin(n\emptyset_i)}{\sum wgt(\emptyset_i) \cos(n\emptyset_i)} \right) / n \quad (1)$$

where n = 1 for the plane, n = 2 for the ellipse, the sum goes over i for the charged particles used in the event plane determinations, $\emptyset_i$ is the azimuthal laboratory angle of particle i, and the quantity wgt will be described below. The angle $\emptyset_1 = \emptyset_{plane}$ covers 0 to $2\pi$ and $\emptyset_2 = \emptyset_{ellipse}$ covers 0 to $\pi$, using the signs of the sums to determine the quadrant. In other words, the angle is defined by the vector whose laboratory components in the plane perpendicular to the beam are, respectively, the numerator and denominator in the above equation. Notice that the angle of the ellipse is calculated by summing over $2\emptyset$ instead of $\emptyset$. Notice also that $p_t$ is not used in these equations, so that they represent the number weighted angles, not the momentum weighted angles. Also, because $p_t$ is not used in the above equations, the acceptance biases have to be removed only with respect to $\emptyset$, and not with respect to $p_t$.

To remove the biases due to acceptance correlations we have used three methods: event plane flattening by weighting, event plane flattening by shifting, and event mixing. Flattening of the event plane laboratory angular distribution by weighting involved using the inverse of the laboratory azimuthal distributions of the particles, summed over all events, as a 36 channel histogram for wgt($\emptyset_i$) in the above equation. Flattening the distribution of the event planes by shifting involved setting wgt($\emptyset_i$) in the above equation to one and then fitting the resultant azimuthal distributions of the event planes, summed over all events, to a Fourier expansion. Harmonics up to fourth order were used for the plane, but of these only the even harmonics entered into the fit for the ellipse. From the resultant coefficients of the fit one can derive an equation for shifting the event plane angles, event by event, to obtain flat distributions.[16] With this method the distributions of $\emptyset_{plane}$ and $\emptyset_{ellipse}$ are flat in the laboratory as shown in the top half of Fig. 1. All the results presented here used this shifting method of flattening even though the flow values and the resolution corrections were exactly the same using the weighting method.

The mixed event method calculates the usual azimuthal distribution of the particles of interest with respect to the event plane of their own event, but then divides this distribution by the azimuthal distribution of these same particles with respect to the event plane of the previous event. This method also gave the same results. In addition, we have obtained the same results by combining the methods: flattening first and then dividing the azimuthal correlations by those for mixed events. However, using the mixed event method with only one mixed event for each real event increases the error of the results by $\sqrt{2}$.

The correlations of the selected particles with respect to the above defined event-by-event planes were found by evaluating the coefficients in the Fourier expansions of the azimuthal distributions (normalized to an average value of one) with respect to the two planes:

$$1 + 2 v_1^{obs} \cos(\emptyset - \emptyset_{plane}) + 2 v_2^{obs} \cos(2(\emptyset - \emptyset_{plane})) \qquad (2)$$

$$1 + \qquad\qquad 2 v_2^{obs} \cos(2(\emptyset - \emptyset_{ellipse})). \qquad (3)$$

The coefficient $v_1^{obs}$ is evaluated by $\langle\cos(\emptyset - \emptyset_{plane})\rangle$, where $\langle\rangle$ indicates the mean value summed over the particles of interest for all events. The $v_1^{obs}$ is $\langle p_x/p_t\rangle$ whereas $v_2^{obs}$ is related to the eccentricity of the ellipse by $\langle(p_x/p_t)^2 - (p_y/p_t)^2\rangle$, with x and y being the directions perpendicular to the beam with x in the event plane. The quantity $v_2^{obs}$ can be evaluated from $\langle\cos(2(\emptyset - \emptyset_{plane}))\rangle$, and in fact, its sign gives the relative orientation of the plane and the ellipse. However, higher accuracy for the value of $v_2^{obs}$ was obtained by evaluating $\langle\cos(2(\emptyset - \emptyset_{ellipse}))\rangle$. Of course, when a particle had been used to calculate the direction of a plane, the auto-correlation effect in its distribution with respect to this plane was removed in the usual way by recalculating that plane's orientation without this particle.[17]

The $v_1^{obs}$ and $v_2^{obs}$ values are the flow values relative to the observed event planes. To obtain the flow values relative to the true reaction plane one has to divide these values by a factor which corrects for the limited resolution of the measurement of the angle of the event planes.[4, 6, 18] To accomplish this the events were randomly divided into two sub-events and the correlations of the planes of the sub-events were evaluated. The square root of $\langle\cos(\emptyset_{sub1} - \emptyset_{sub2})\rangle$ and of $\langle\cos(2(\emptyset_{sub1} - \emptyset_{sub2}))\rangle$ are the resolution corrections of the observed sub-event planes. The resolution corrections of the observed event planes of the full events were determined by correcting for the fact that the full events have twice the multiplicity of the sub-events. When the resolution corrections are small compared to one this can be done by multiplying the resolution corrections by $\sqrt{2}$. Instead we used the more general multiplicity dependence of the resolution correction given by eq. 13 and fig. 4 of ref. [4] to do this extrapolation. Table I lists the measured resolution correction factors for the sub-events as well as the extrapolated values for the full events.

Table I. Corrections for the
resolutions of the observed planes.

|  | sub-event | full event |
|---|---|---|
| plane | 0.25 ± 0.006 | 0.35 ± 0.009 |
| ellipse | 0.19 ± 0.008 | 0.27 ± 0.011 |

To evaluate our methods and, in particular, our ability to remove the acceptance correlations, we generated 50k events with a simple Monte-Carlo event generator which had no azimuthal correlations but reproduced the charged particle and proton multiplicities and $p_t$ spectra within our acceptance. These events were filtered through a GEANT model of the NA49 detector.[19] In the GEANT simulation, all physics processes including decays were turned off so that these events should not have any correlations beyond those due to the acceptance geometry.

The experimental azimuthal distributions of the charged particles are plotted with respect to the charged particle plane and ellipse in Fig. 1 bottom. Also shown are the results from the simple Monte-Carlo filtered for the NA49 acceptance. The graphs clearly show both directed flow in the forward hemisphere (with symmetry about 180°) and elliptic flow (with symmetry about 90°). In the bottom left graph the fit contains a second harmonic with positive amplitude which shows that the ellipse is aligned with the plane. This means that the elliptic flow is in-plane, not out-of-plane squeeze-out. This was verified by observing a positive correlation between the plane of one charged particle sub-event and the ellipse of the other.

Data not shown indicate that the ellipses of the protons and the other charged particles are aligned, and that the directed flow of the protons appears to be small and opposite to that of the other charged particles. It was assumed that the proton directed flow is in the positive (direction of the impact parameter from the target nucleus) side of the event plane. A summary of the results is given in Table II, integrated over the rapidity and $p_t$ ranges indicated. This selection of charged particles is the same as that which was used to determine the event planes. The units of percent mean that the numbers have been multiplied by a factor of one hundred.

Table II. Flow values integrated over the indicated y and $p_t$ ranges.

| Particle | y | $p_t$ (GeV/c) | $v_1$ (%) | $v_2$ (%) |
|---|---|---|---|---|
| protons | 3 - 6 | 0.0 - 2.0 | 1.1 ± 0.2 | 2.6 ± 0.3 |
| charged particles | 4 - 6 | 0.05 - 1.0 | -3.0 ± 0.1 | 2.3 ± 0.1 |

The rapidity and $p_t$ dependence of the flow of the protons and the identified pions are now evaluated relative to the same charged particle event planes used above. As above, all flow values are corrected for the resolutions of these planes. For the rapidity dependence a rather high $p_t$ window of 0.6 to 2.0 GeV/c was used for the protons but a low window of 0.05 to 0.35

GeV/c was used for the pions. These windows were chosen to obtain the widest region of flat acceptance. The rapidity dependence of the flow is shown in Fig. 2 with reflection about mid-rapidity. In reflection, the signs of the $v_1$ values have been reversed in the backward hemisphere, but not the $v_2$ values. The directed flow ($v_1$) values exhibit characteristic S-shaped curves and the elliptic flow ($v_2$) values appear to be flat for the pions but peaking somewhat near mid-rapidity for the protons. Here one can see that our choice of the sign of the event plane is plausible from the fact that the protons at high rapidity have positive directed flow, as one would expect for the baryons in the reaction plane. Fig. 3 shows the $p_t$ dependence of the flow in a rapidity window from 4.0 to 5.0. These curves should go to zero at zero $p_t$ where no transverse direction is defined. For the pions the lowest points indicate that the curves are tending to zero.

At first sight the $v_1$ curves in Fig. 3 appear peculiar, especially for the pions, because they approach zero from the negative side. However, this behavior was predicted for protons by Voloshin[20] as a consequence of the interaction of transverse radial flow and directed flow. Simply, in the presence of large transverse radial flow, a low $p_t$ particle can be produced only by the part of the moving source where the directed flow subtracts from the radial flow. If this is the correct explanation then the data also contain information on the transverse radial flow. However, especially for the pions, it is also possible that this behavior of the directed flow results from some kind of fireball shadowing effect, resonance decays, or Coulomb effects.

Previously, elliptic flow has been observed using the NA49 Ring Calorimeter by analyzing the azimuthal anisotropy of the transverse energy in the pseudorapidity interval from 2 to 4 for an impact parameter of 7 to 8 fm.[21] The correlation which was observed between the forward and backward event planes is not inconsistent with that seen here considering that transverse energy flow, not number flow, was studied, that neutral particles and charged baryons were included, and that the bins were in pseudorapidity. Also the elliptic flow of photons from $\pi^0$ decay has been reported[22] by WA93 for S + Au at the SPS. They find an anisotropy of the order of 5% for semi-central collisions.

At the AGS, E877 reported[18] $v_1$ values of about 10% for protons and about 2% for pions. Their $v_2$ values[13] for charged particles are, however, at most 2%. Thus, although the directed flow is smaller at the SPS, the elliptic flow may be larger.

In summary, we have presented the first data on directed and elliptic flow for Pb + Pb collisions at 158 GeV/nucleon. Protons and pions exhibit significant, but opposite, directed flow at large rapidities. The elliptic flow signal was found to be fairly independent of rapidity for the pions but peaking somewhat near mid-rapidity for the protons. For both sets of particles the flow axis of the elliptic flow is in the plane of the directed flow. This excludes shadowing by spectator matter as the origin of the elliptic flow. Therefore we conclude that the elliptic flow in these semi-central collisions retains some signature of the pressure in the high density region created during the initial collision.

*Acknowledgments*. This work was supported by the Director, Office of Energy Research, Division of Nuclear Physics of the Office of High Energy and Nuclear Physics of the US Department of Energy under Contract DE-ACO3-76SFOOO98, the US National Science Foundation, the Bundesministerium fur Bildung und Forschung, Germany, the Alexander von Humboldt Foundation, the UK Engineering and Physical Sciences Research Council, the Polish State Committee for Scientific Research (2 P03B 01912), and the Polish-German Foundation.

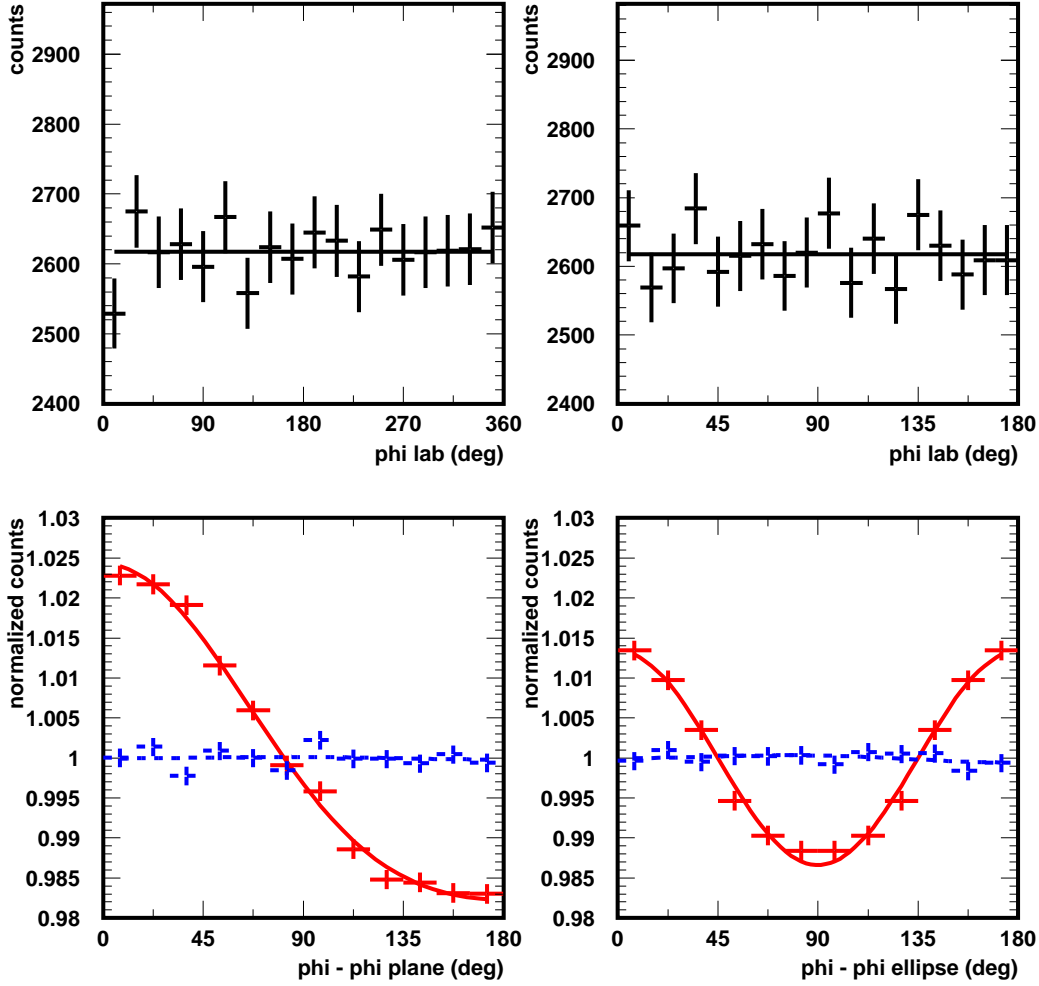

Fig 1. On the top are shown the flatness achieved with the shifting method for the plane (left) and ellipse (right) in the laboratory system. Note the suppressed zero on the vertical axis.

On the bottom are shown the azimuthal distributions of the charged particles with respect to the plane (left) and ellipse (right) of the charged particles. The distributions have been normalized to an average value of one. The dashed points and curves near a value of one are for the simple Monte-Carlo. The curves are fits with $\cos(\emptyset)$ plus $\cos(2\emptyset)$ (eq. 2) (left), and $\cos(2\emptyset)$ (eq. 3) (right). On the right, the points above 90° have been reflected from those below 90°. The results are integrated for rapidity from 4.0 to 6.0 and $p_t$ from 0.05 to 1.0 GeV/c.

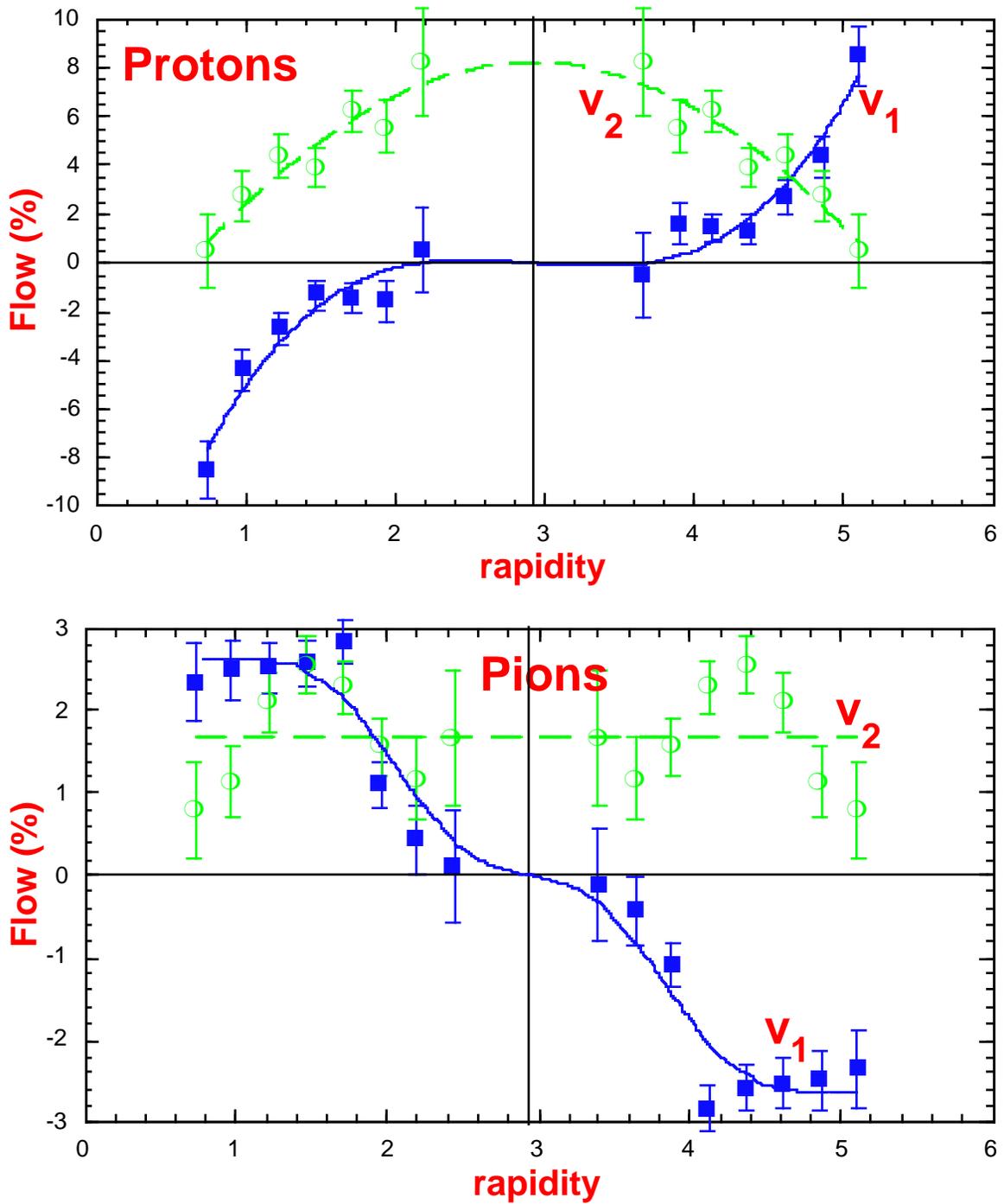

Fig 2. The rapidity dependence of the directed ($v_1$) and elliptic ($v_2$) flow for the protons (0.6 < $p_t$ < 2.0 GeV/c) and pions (0.05 < $p_t$ < 0.35 GeV/c). The points between y=0 and mid-rapidity (y=2.92) have been reflected from the measurements in the forward hemisphere. The curves are to guide the eye.

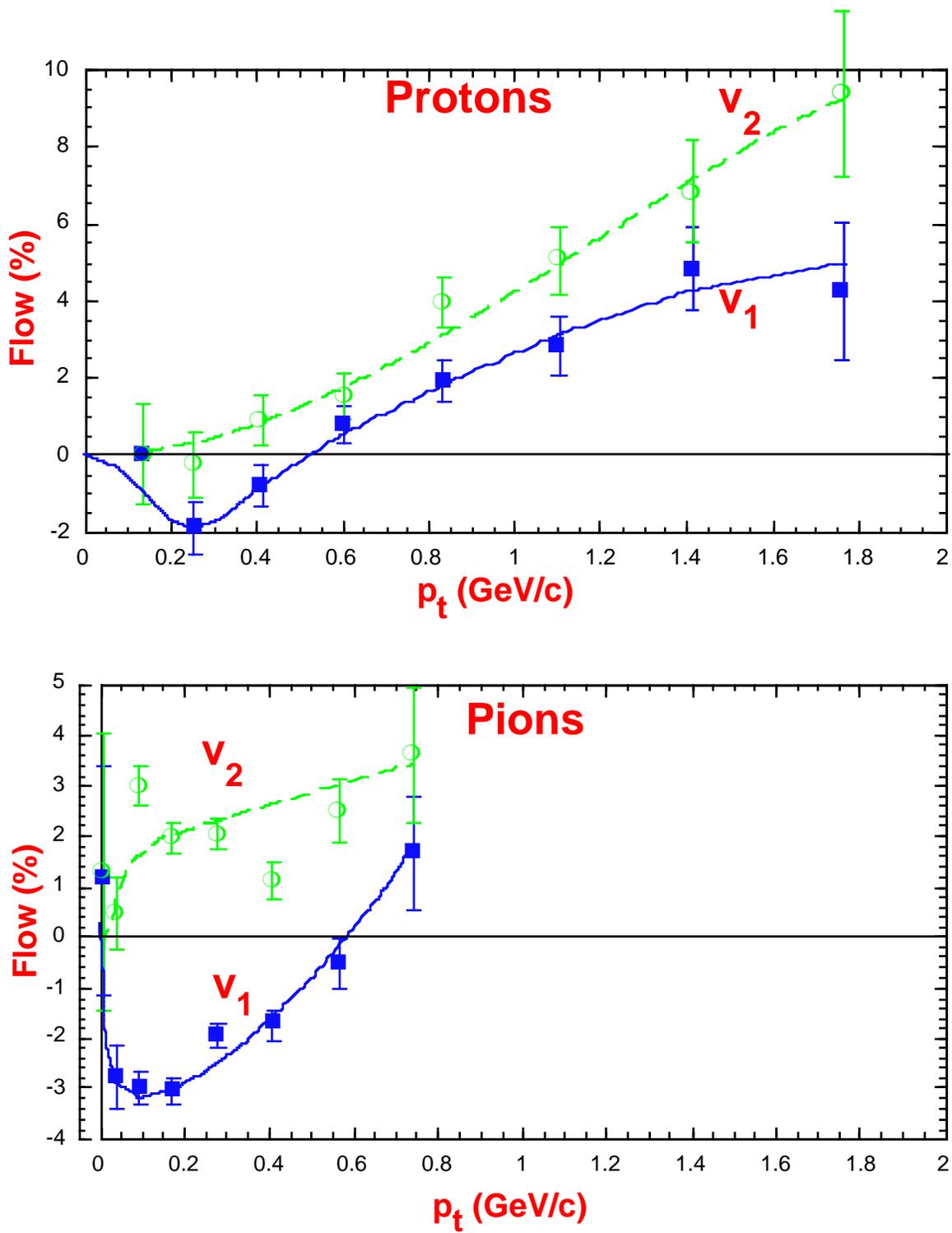

Fig 3. The transverse momentum dependence of the directed ($v_1$) and elliptic ($v_2$) flow for the protons and pions with $4.0 < y < 5.0$. The curves are to guide the eye.

± deceased.
+ present address: Max-Planck-Institut für Physik, Munich, Germany.
∗ present address: Kent State Univ., Kent, OH, USA.
# present affiliation: Physikalisches Institut, Universität Heidelberg, Germany.
& permanent affiliation: Moscow Engineering Physics Institute, Moscow, Russia.